# Privacy-preserving data outsourcing in the cloud via semantic data splitting

## David Sánchez[1]

*UNESCO Chair in Data Privacy, Department of Computer Science and Mathematics, Universitat Rovira i Virgili, Avda. Països Catalans, 26, 43007 Tarragona (Spain)*

## Montserrat Batet

*Internet Interdisciplinary Institute (IN3), Universitat Oberta de Catalunya, Avda. Carl Friedrich Gauss, 5, 08860 Castelldefels, Barcelona (Spain)*

## Abstract

Even though cloud computing provides many intrinsic benefits (e.g., cost savings, availability, scalability, etc.), privacy concerns related to the lack of control over the storage and management of the outsourced (confidential) data still prevent many customers from migrating to the cloud. In this respect, several privacy-protection mechanisms based on a prior encryption of the data to be outsourced have been proposed. Data encryption offers robust security, but at the cost of hampering the efficiency of the service and limiting the functionalities that can be applied over the (encrypted) data stored on cloud premises. Because both efficiency and functionality are crucial advantages of cloud computing, especially in SaaS, in this paper we aim at retaining them by proposing a privacy-protection mechanism that relies on splitting (clear) data, and on the distributed storage offered by the increasingly popular notion of *multi-clouds*. Specifically, we propose a *semantically-grounded data splitting mechanism* that is able to automatically detect pieces of data that may cause privacy risks and split them on local premises, so that each chunk does not incur in those risks; then, chunks of *clear* data are independently stored into the separate locations of a multi-cloud, so that external entities (cloud service providers and attackers) cannot have access to the whole confidential data. Because partial data are stored in clear on cloud premises, outsourced functionalities are seamlessly and efficiently supported by just broadcasting queries to the different cloud locations. To enforce a robust privacy notion, our proposal relies on a privacy model that offers *a priori* privacy guarantees; to ensure its feasibility, we have designed heuristic algorithms that minimize the number of cloud storage locations we need; to show its potential and generality, we have applied it to the least structured and most challenging data type: plain textual documents.

---

[1] Corresponding author. Address: Departament d'Enginyeria Informàtica i Matemàtiques. Universitat Rovira i Virgili. Avda. Països Catalans, 26. 43007. Tarragona. Spain

Tel.: +034 977 559657; Fax: +034 977 559710;

E-mail: david.sanchez@urv.cat.

# 1. Introduction

Cloud computing offers many benefits for companies, public institutions and individuals willing to store and process their data in the cloud, such as, dynamically scalable resources, improved agility and manageability, scalability, availability and universal data access independently of geographical location, thus providing computational power and flexibility. Most important, cloud computing generally implies costs savings since it reduces infrastructure and maintenance costs, thus providing cheap storage capacity and computing. In relation to this, IBM forecasted 60% of the big organizations are ready to embrace cloud computing in the next years [1].

However, security concerns about data loss or leakage since the lack of direct control over the storage and management of the outsourced data have proved to be a real threat that prevents many customers from migrating to the cloud. In the EU, 39 % of enterprises (57% for large ones) using the cloud reported the risk of a security breach as the main limiting factor in the use of cloud computing services. More specifically, in a report by the cloud Security Alliance over 165 IT and security professionals in the U.S., most of the respondents considered cloud storage as high risk [2]. The European Network and Information Security Agency identified "loss of governance" over the data outsourced to the cloud as a critically important factor [3]. As an example, Dropbox is supposed to encrypt user data with the aid of heavy cryptography, but private keys are created and managed by Dropbox itself and not by the data owner. Moreover, recent security breaches compromising the data of millions of people have mined the trust of potential users in the cloud. Some well-known examples include the Sony PlayStation Network outage[2] as a result of an external intrusion, in which personal details from approximately 77 million accounts were stolen, the multi-day outage in Dropbox[3] that temporarily allowed visitors to log into any of its 25 million customer accounts as a result of a misconfiguration problem, or the leakage of private photos of a number of celebrities from the Apple iCloud storage service as a result of weakly protected login credentials[4].

From another perspective, cloud users may have concerns about what Cloud Service Providers (CSPs) intend to do with their (potentially confidential) data. Cloud computing has given CSPs the opportunity to analyze and exploit large amounts of personal data. For example, a recent privacy policy in Google[5] specifies that whatever information a user decides to share through any Google service can be used, reproduced, modified or distributed by Google with the aim of improving or promoting its services (e.g., the Gmail filtering system scans the content of our emails to serve personalized ads). In a similar way, Yahoo! also reserves the right to collect and use users' data and combine them with information obtained from business partners[6]. Data collected by CSPs can be used to benefit the users (e.g., by providing personalized services) but, at the same time, they may raise privacy concerns. In a report of the Federal Trade Commission (FTC), it is stated that CSPs regularly collect and analyze users' data without users' knowledge, and that some of the inferences of that analysis could be sensitive; for example, a CSP could identify the users that have diabetes because of their interest in sugar-free products and share this information with an insurance company that could use this information to classify a person as higher risk [4]. In the current socioeconomic context in which electronic data are often considered "the new oil" of this century, user concerns about the data monetization actions performed by CSPs and about the harmful consequences that this may have (e.g., discriminatory actions in job applications, medical insurances, etc.) are perfectly understandable.

To mitigate these problems and to regain the user's control over the protection applied to her confidential data prior outsourcing them to the cloud, several mechanisms have been proposed. They mainly apply a certain kind of data protection on the client side so that only protected outcomes are outsourced to the cloud and so that only the data owner is able to correctly reconstruct the data retrieved from the cloud. We discuss them in the next section.

## 1.1 Related work

Encryption is a natural solution for enforcing data protection in the cloud [5]. Several solutions based on public and symmetric key cryptography have been presented for cloud storage in which data protection is provided via encryption. Encryption is performed before the data are transmitted and stored in the cloud, being decrypted only after having been returned to the data owner. These solutions have been usually enforced as trusted encryption proxies. For example, CipherCloud provides a secure gateway located in a trusted environment (i.e., the user's local network) targeted at several popular Software-as-a-Service CSPs (e.g., Gmail, Salesforce, Amazon, etc.). It applies encryption to specific user data (e.g., mail bodies and subject, chat messages, etc.) before storing them in the cloud. Cloud services are replicated in the secure gateways (i.e., web interfaces, business logics) to provide coherent results to operations like searching or sorting. Encryption keys are managed and stored locally under the control of the user. PerspecSys follows a similar approach, which consists of a server and a reverse proxy that implement data encryption methods. SecureCloud is a solution for Infrastructure-as-a-Service cloud-based execution platforms in which application data are stored encrypted in the cloud, leaving the key control and the definition of access policies to the user. Both BoxCryptor and Cloudfogger focus on providing client-side data encryption for cloud storage services like Dropbox or Google drive, by using basic symmetric key encryption (i.e., AES).

Even though the above systems offer robust privacy-preserving data storage for the end users, they face several problems, either technical or user-oriented:

- The communication between the CSP and the client device is captured and "reverse-engineered" in order to enrich it with security features (data encryption) that are transparent to both the CSP and the client. However, the communication protocol may often change, which would require a continuous update of the provided security features, a difficult task that could seriously impair the reliability and availability of the service. In addition, the CSP itself might also implement specific measures to prevent this approach if it is uncomfortable with users that systematically upload encrypted contents to its servers. This is especially relevant for CSPs that offer their services free of charge because they expect to gain profits from the analysis of users' data (e.g., Google have legal possibilities to access these data, due to the terms of service agreement) and which may ban users that only provide encrypted –and thus, useless- data.

- Because sensitive data are systematically encrypted and stored at the CSP, which is supposed to be unaware of this fact, functionalities offered by the CSP can yield gibberish outcomes. In this case, only a reduced set of functionalities can be preserved (basically, data storage and plain retrieval), or encryption can be applied only on data that are not processed in the cloud (e.g., binary files), or cloud services must be replicated at the trusted gateway. In the latter case, the gateway is forced to redundantly store unencrypted data and to re-implement and reverse-engineer some cloud services, thus defeating the whole purpose of data and computation outsourcing. Even though in recent years some cryptographic solutions have been proposed with a limited support for a number of operations over encrypted data (mainly searches), complex operations would require from solutions like homomorphic encryption, which are still far from being efficiently applicable in a real setting [6]. Even the more efficient searchable encryption solutions [7-9] still require adding a considerable amount of data (e.g., hierarchical indexes) to the outsourced data, performing several queries to retrieve the matched data and/or offer a limited support for complex conjunctive queries involving logical (AND/OR) and relational operators (><) and value ranges.

- Encrypting the whole data uploaded to the CSP at the client side implies the loss of several degrees of magnitude in efficiency with regard to both storage and processing, which in the case of cloud computing it would mean defeating its own purpose, because one of the main motivation for moving to the cloud, in addition to the provided functionalities, is saving costs [10]. Moreover, the management of encryption keys may add new security risks at the client side.
- The vast majority of users are not familiar with the fundamental concepts of cryptography and many of them are not capable of properly managing keys or certificates [11]; thus, the effectiveness and security of cryptographic solutions may be compromised because of a negligent management of cryptographic materials.

To circumvent these problems, we propose using privacy-preserving solutions alternative to data encryption, which manage data in clear form. To enforce these, we will rely on the increasingly popular notion of *multi-cloud*, that is, the use of multiple cloud computing services in a single heterogeneous architecture [12]. Multi-clouds add several advantages, such as reducing reliance on any single vendor, increasing flexibility or mitigating disasters, but we are specifically interested in the distributed and unconnected nature of multi-cloud services. This opens the door for alternative data protection techniques based on data *partitioning* or *splitting* [13].

With data splitting, sensitive data in a document or in a data base are split in chunks and stored (in clear form) in different locations so that individual parts do not disclose identities and/or reveal confidential information. Specifically, the distributed storage of chunks reduces the amount of information that can be gathered by a third party, thus rendering the knowledge acquired from data analysis or user profiling incomplete and ambiguous, and also minimizes the consequences of potential leakages [14]. At the same time, the fact that (partial) data are stored in clear form makes it possible to seamlessly retain a number of cloud functionalities (e.g., keyword searches, conjunctive and range queries, statistical calculations, etc.) by just broadcasting user queries to the different locations. Even though these functionalities provide partial results because they have been applied in partial data chunks, it is possible for the data owner, who knows the location of each chunk, to reconstruct the complete result by combining and aggregating partial results [15], in an operation much more efficient than data decryption. Thus, in comparison with data encryption and, in addition to supporting a larger variety of functionalities, data splitting does not have such large overhead associated with query processing [14, 15]. Finally, in scenarios in which the CSP expects non-encrypted data, individual chunks (e.g., pieces of a document, separate attributes value of a database) may still be useful, because the CSP can still perform data analyses and extract -partial- conclusions.

Many of the splitting mechanisms proposed in the literature perform the data partitioning at a binary level. For example, in [16], the authors split user files in bytes, which are shifted and recombined into a fixed set of parts which are finally stored in different locations. In [14], each file to be outsourced is associated with a privacy level, which is chosen by the user according to the sensitivity of its contents. Then, chunks of files are created based on the standard RAID storage mechanism, and those with a higher privacy level are stored in the location in which the user has more trust. This kind of mechanisms, in which data partition is done according to a fixed criterion and bytes are alternatively picked and recombined, are well-suited for purely binary or even multimedia files, whose contents are usually stored but not processed by the cloud. However, for files whose contents should be processed by the cloud (e.g., textual documents such as e-mails or Word files, structured databases, etc.), in order to provide functionalities (e.g., searches, computations), the output they produce offers neither privacy nor utility guarantees. On the one hand, it is not possible to ensure that chunks of bytes do not contain enough data to cause disclosure if they are too large; on the other hand, it is not possible to guarantee cloud functionalities or retain partial data utility for the CSP if chunks are too small or arbitrarily combined. This makes it difficult for the users to understand the kind of protection and utility preservation that a specific method is achieving, which inherently depend on the semantics of the data to be protected.

At an attribute level, all data splitting approaches focus on structured (SQL) databases. In this scenario, data splitting can be horizontal (sets of data records are stored separately) or vertical (sets of attributes are stored separately). However, as stated in [15], horizontal partitioning is of limited use in enabling privacy-preserving decomposition because disclosure often arises from the correlation between attributes rather than from different records, which are usually independent. Vertical splitting is proposed in [13, 15, 17, 18] to ensure confidentiality at the record level: identifying attributes of the database (i.e., those that unequivocally identify the individual assocaited to the record, such as the SS number) are encrypted, whereas quasi-identifiers (i.e., those that individually do not cause disclosure but that, in combination, may unequivocally identify individuals, such as job+address+age), are split and stored in different locations. Both [15] and [18] limit the number of locations to two, because they assume that there is a unique set of quasi-identifying attributes. which does not necessary hold. In this manner, disclosure is avoided while maintaining data utility at attribute level. Moreover, they also propose different algorithms to determine the best decomposition for minimizing the cost of the subsequent SQL queries, which need to be broadcasted and their results aggregated. Even though these methods offer more robust privacy guarantees and better preserve functionalities than binary appraoches, they strictly rely on the structure of the data (i.e., attributes and records) and on a set of manually defined rules that specify the combinations of attributes that may cause disclosure and, thus, the splitting criterion. As a result, they cannot be applied to unstructured data, such as documents, and may not scale well with large datasets because of the human intervention needed to define the risky combination of attributes for each dataset. Moreover, it has been acknowledged that defining such risky combinations is a costly and difficult task even for expert human sanitizers [19].

## 1.2   Contributions and plan

To tackle the limitations of the related works depicted above, in this paper we present a semantically-grounded data splitting protection mechanism (*semantic data splitting* for short) for data outsourced to the cloud. As the main innovation, our approach assesses the *actual semantics* of the data to be protected and of the privacy requirements of the user in order to i) automatically detect the pieces of data that may cause disclosure and ii) automatically orchestrate the splitting of data so that the privacy of the user is guaranteed. Contrary to methods partitioning data at a binary level [14, 16], in our approach, the data partition and the distributed storage of data chunks are driven by the semantics and sensitiveness of the data contents and the privacy requirements of the user. Thus, the contents of the resulting chunks are not arbitrary, and it is ensured that they do not incur in privacy breaches. This also contrasts with approaches in which the responsibility of defining the specific data to be protected (e.g., files, DB attributes, etc.) is left to the user [14, 18], a solution that may be unfeasible in many scenarios, given the large amount of documents that are outsourced to the cloud and the lack of knowledge of many users regarding the privacy risks of their own data. Finally, contrary to methods focusing on structured databases [15, 17, 18], our approach is solely based on data semantics, and thus, it can be applied to any kind of data, regardless of whether they are structured or not. To show the potential and the generality of our approach, in this paper we focus on the storage of raw textual documents (e.g., e-mails, Word files, etc.), which are the most challenging to protect because of their lack of structure [19] while, at the same time, they constitute the most common way to exchange -potentially sensitive- information between human actors.

The semantic ground of our method is inspired by the state of the art in document sanitization/redacting techniques [20-24], which strive at automatically identifying and protecting sensitive entities in input documents. To provide *a priori* privacy guarantees, our proposal is designed to fulfill an inherently semantic privacy model for document protection: *C*-sanitization [25, 26]. By instantiating and enforcing this privacy model, data owners can intuitively define their privacy requirements according to the semantics of the data that they do not want to disclose to the cloud and to potential attackers.

The rest of this paper is organized as follows. Section 2 discusses the security model and presents the architecture of our system. Section 3 discusses the notions related to the protection of plain unstructured documents and the privacy model in which our approach relies. Section 4 presents the privacy-preserving methodology that detects risky terms and splits and stores document chunks in a multi-cloud while fulfilling the privacy requirements of the user. Moreover, we also present some heuristics to minimize the number of cloud locations we need to fulfill with those requirements, and detail how functionalities (retrieval, updates, and searches) can be seamlessly supported and how data and query results can be semantically reconstructed. Section 5 discusses the security, efficiency and support of outsourced functionalities of our approach against cryptographic methods. Section 6 details the evaluation of the system with several highly sensitive documents and privacy criteria, and measures the efficiency of the allocation of data chunks according to a cost metric. The final section presents the conclusions and depicts several lines of future research.

## 2. Security model and system architecture

We assume in our security model that the user does not trust any CSP to properly preserve her privacy. We also assume that all CSPs are *honest but can be curious*, that is, they may actively gather, compile and analyze the data they store, as well as the queries they receive from the users if, by doing so, they expect to gain additional user data or break any privacy measure that the user may have implemented over the data prior outsourcing them. In any case, the CSP, as a commercial provider of services, will not act maliciously against the user (e.g., consciously altering or deleting data) and will follow the protocols as expected, i.e., it is *honest*.

We also assume the availability of several CSPs that provide a homogenous service, that is, a *multi-cloud*, and offer a pool of locations in which data can be stored. In principle, we assume that the different CSPs are independent from each other and, thus, they do not collude to aggregate the partial data they store in the hope of breaching the user privacy. In any case, as we discuss below, our system will also offer some intrinsic protection in the advent of such collusion.

Finally, we assume that the user side is fully trusted and secure, because it was the place in which the data were initially stored in clear form. Because of this, we can deploy a local application or a proxy on the user side that can securely process the clear data prior outsourcing it to the cloud and store the metadata needed to process subsequent queries on those data and reconstruct the results. In some scenarios (e.g., an enterprise intranet), this secure application/proxy could be shared among different end users (e.g., employees). In such case, cloud storage locations will be shared among the different users. The application/proxy will manage the pool of available cloud locations transparently for the users.

Moreover, the user may specify a set of privacy requirements, which define the topics that should be protected in any data outsourced to the cloud (e.g., identities, locations, healthcare data, etc.), and up to what degree, i.e., the maximum level of semantic disclosure that the user allows for each topic. Requirements may be related to identities, thus requiring protection against *identity disclosure*, and/or to confidential data, i.e., *attribute disclosure*. Contrary to previous works [15, 18], these requirements are defined at a conceptual level rather than at data/structural level (e.g., combinations of risky attributes in a structured database). According to the scenario described above, the architecture we propose is depicted in Figure 1.

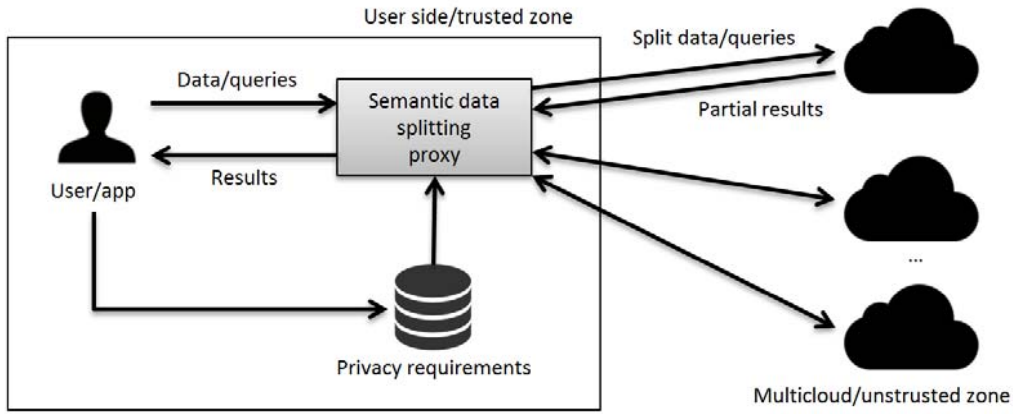

*Figure 1.* General architecture of the system.

The semantic data splitting proxy performs two main actions: data storage in the cloud and data retrieval from the cloud as a result of a user query. The logics of the first action are depicted in Figure 2. First, the proxy receives the data to be stored and assesses the disclosure risks of the content. To do so, it relies on the privacy requirements defined by the user, so that those terms whose presence or co-occurrence in the input data may violate the user privacy are automatically tagged as *risky*. These tagged data are then passed to the splitting module that, according to the risky terms and the constraints defined in the privacy requirements, decides how data are split and how many storage locations are needed; each piece of data that can be safely stored together constitutes a *data chunk*. It also stores the splitting criterion in a local database as metadata, so that the system can properly process future queries over the data and aggregate the partial results, and forwards each data chunk to a separate CSP. In this manner, we ensure that the splitting process is *lossless*, that is, we can reconstruct the original results if needed without having to store the whole data on local premises, and *privacy preserving,* since the contents stored in each separate CSP do not incur in a privacy threat according to the privacy requirements of the user [15].

Notice that, if several users employ the same proxy and, thus, share the same pool of cloud locations, each location may store chunks of clear data from different users, i.e., the *split* data of several users are *merged* at each cloud location. Because the proxy is the only one capable of interpreting which chunk corresponds to which user, this provides a degree of protection against the collusion of several CSPs: even though CSPs may orchestrate efforts to aggregate the partial data they store, they will fail because they are not able to unequivocally discriminate the user to which each data chunk corresponds.

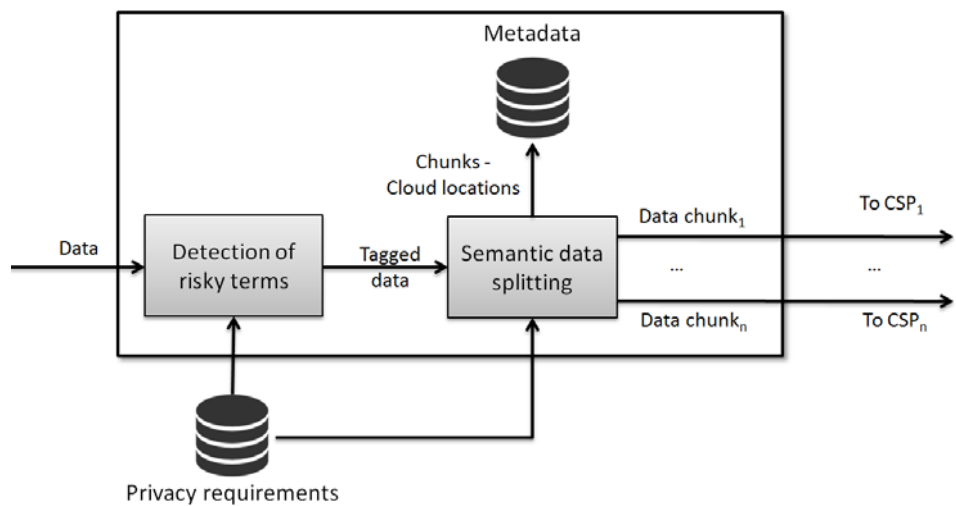

*Figure 2.* Data outsourcing workflow to a multi-cloud via semantic data splitting.

On the other hand, the logics related to the execution of a user query over the outsourced (split) data and the aggregation of the final results are depicted in Figure 3. First, the query is processed, so that the storage locations of the data the query refers to are retrieved from the local metadata database. Then, the query is replicated as many times as data chunks the data have been split into, and these queries are forwarded to the corresponding CSPs. As a result, each CSP provides a set of results that represent a partial view of the complete result. Finally, the proxy aggregates these results according to the criterion used to split them, also stored as metadata, and provides the complete result to the user.

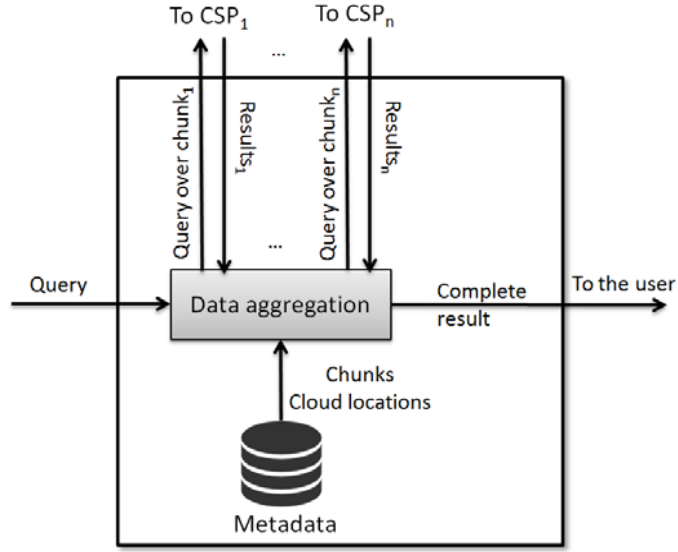

*Figure 3.* Query answering workflow with split data.

## 3.   Data protection with privacy guarantees

Several privacy models have been proposed by the computer science research community to offer sound privacy guarantees to data protection. Many of these, such as the well-known *k*-anonymity model [27, 28] and its extensions to plain textual documents [29], exploit the regular structure of the data (e.g., the records in a structured database, collections of documents with an homogenous structure) to define the privacy guarantee as a probability of re-identification of one individual within the dataset to be released. Privacy is thus enforced by making the individual's identifying data indistinguishable enough within the dataset, i.e., it mainly protects data against *identity disclosure*. Other well-known models, such as differential privacy [30], also deal with structured data sets and require attributes with bounded domains.

However, in our setting, documents are generated independently by each user and, thus, they should be protected individually according to the semantics they disclose; free text is also unstructured and unbounded. Moreover, we would like the instantiation of the privacy model to be intuitive enough so that it can be employed by users to clearly define their privacy requirements; at this respect, most privacy models are instantiated by means of abstract numerical values [21, 28, 30], which many users find difficult to understand [31, 32]. Finally, we would also like to seamlessly support protection against *identity* and/or *attribute* disclosure.

As far as we know, the only privacy model that fits with this scenario and these requirements is *C*-sanitization [25, 26], a general privacy model for (textual) document sanitization. *C*-sanitization defines the desired level of privacy by means of a set of topics, *C* (e.g., identities, confidential data) in a way that the privacy guarantees are fulfilled if the protected outcome does not contain any term that, individually or in aggregation with other terms co-occurring in the same document, reveals the semantics of the sensitive entities in *C*. For example, given a medical record, its {*AIDS*, *HIV*}-sanitized version should not contain terms (e.g., synonyms or acronyms of AIDS and HIV, but also combinations of treatments or drugs) by which a knowledgeable attacker can unequivocally infer that the protected document contents refer to AIDS or HIV.

Formally, the model is defined as follows:

**Definition 1**. (*C-sanitization*). Given an input document $D$, the background knowledge $K$ that is available for potential attackers, and a set of sensitive entities $C$ to be protected, we say that $D'$ is the *C-sanitized* version of $D$ if $D'$ does not contain any term $t$ or group of terms $T$ that can, individually or in aggregate, unequivocally disclose *any* entity in $C$ by exploiting $K$.

To more accurately configure the tradeoff between the degree of protection (i.e., prevention of sensitive information disclosure) and data utility resulting from protection, we can define a threshold of maximum disclosure by specifying a set of generalizations $g(C)$ for the sensitive entities in $C$ [26]. For example, an {(*AIDS*, *disease*), (*HIV*, *virus*)}-sanitized document would not only prevent disclosing AIDS and HIV but, at most, it would only disclose that the contents refer to a (generic) disease and virus, respectively. Formally:

**Definition 2**. ((*C, g(C))-sanitization)* Given an input document $D$, the background knowledge $K$, an ordered set of sensitive entities $C$ to be protected and an ordered set of their corresponding generalizations $g(C)$, we say that $D'$ is the (*C, g(C))-sanitized* version of $D$ if $D'$ does not contain any term $t$ or group of terms $T$ that individually or in aggregate and, for all $c$ in $C$, disclose more than the semantics provided by their respective $g(c)$ by exploiting $K$.

By instantiating the *C*-sanitization model, users can intuitively define their privacy requirements, because the parameter $C$ states, by means of linguistic labels, the set of sensitive topics that the protected document would not disclose. Another advantage of this model is that it can be easily instantiated to enforce the privacy requirements of current legislations and regulations on data privacy [33-38], whose definitions of sensitive data are also linguistic (e.g., religion, sexuality, race, healthcare data, etc.). Finally, the use of generalizations, $g(C)$, as disclosure thresholds for $C$, permits to lower the amount of semantics that can be disclosed about $C$, thus forcing the system to implement a stricter sanitization. Semantically, the use of generalizations in the model instantiation to finely tune the privacy guarantees is very intuitive and provides a clear picture on the amount of knowledge that external entities (readers, data analysts and attackers) can learn of $C$ in the protected document.

# 4. Data outsourcing with privacy guarantees via *C*-sanitization and semantic data splitting

In this section, we detail our semantic data splitting mechanism that, on the contrary to related works on data splitting [13, 15, 18], is able to automatically orchestrate the splitting process, and the subsequent queries and data retrieval, according to the semantics and privacy risks inherent to the data. To offer privacy guarantees, the protection offered by our solution fulfills the *C*-sanitization model introduced above; in this way, the partition and distributed storage of chunks in separate CSPs will be done according to the semantics and sensitiveness of the document contents and the privacy requirements of the user (stated via instantiating the model), in a way that the chunks separately stored in the cloud do not incur in privacy breaches.

The process consists of two steps: (i) detection of those terms in the input document that cause disclosure risk according to the privacy requirements; and (ii) splitting and distributed storage of those terms in order to prevent disclosure.

## 4.1. Detection of risky terms

The automatic detection of risky terms occurring (and co-occurring) in the input document is driven by the notion of disclosure stated by the instantiation of the *C*-sanitization model, which states the privacy requirements. In this respect, the semantic disclosure that terms $t/T$ cause with regard to the entities to be protected *C,* can be naturally enforced in information theoretic terms by quantifying the amount of information given by $t/T$ about *C*. Under the umbrella of the Information Theory, the semantics encompassed by an entity *c* can be quantified according to its Information Content ($IC(c)$); likewise, the semantics that a term *t* or a group of terms *T* discloses about an entity *c* are measured according to their Point-wise Mutual Information ($PMI(c,t)$). Thus, we can reformulate the general Definition 1 in information theoretic terms as follows:

**Definition 3**. (*Information Theoretic C-sanitization*). Given an input document *D,* the background knowledge *K* and a set of sensitive entities *C* to be protected, we say that *D'* is the *C-sanitized* version of *D* if, for all *c* in *C, D'* does not contain any term *t* or group of terms *T* so that, according to *K, PMI(c;t)=IC(c)* or *PMI(c;T)=IC(c)*, respectively.

By applying this to our scenario, those terms *t* (or group of terms *T*) in the input document whose *PMI(c;t)=IC(c)* (or *PMI(c;T)=IC(c)*) for any *c* in *C* will cause *disclosure risk*. The *IC* of the entity *c* measures the information/semantics that should be hidden because of its sensitive nature, while the PMI measures the amount of information/semantics that a term *t* or group of terms *T* reveal about *c*. Figure 4 illustrates the relationship between the *IC* and the *PMI* of two entities: *cancer* is the entity *c* to be protected and *surgery* is the term *t* occurring in the document, which is semantically related to *c*. The figure also highlights, in gray, the disclosure of information/semantics that *t* causes to *c*.

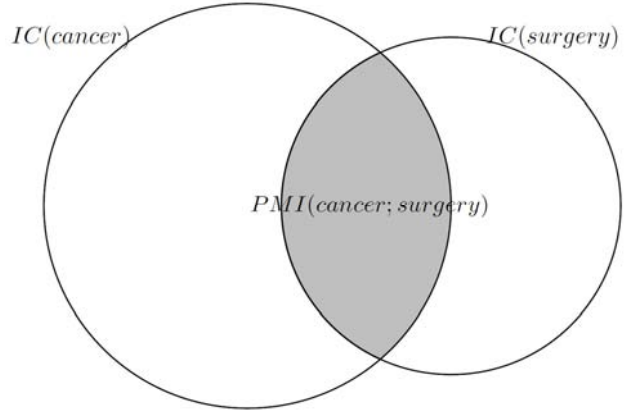

*Figure 4*. Left circle: information/semantics encompassed by *cancer* (the entity *c* to be protected); right circle: information/semantics encompassed by *surgery* (the term *t* appearing in the input document); grayed area: information/semantics that *surgery* discloses from *cancer*.

The disclosure risk assessment can be extended (and made stricter) if generalizations of the sensitive entities *g(C)* are defined as thresholds of maximum disclosure. By adapting Definition 2 in information theoretic terms, disclosure risk would now happen for those terms *t* (or group of terms *T*) in the input document whose *PMI(c;t)>IC(g(c))* (or *PMI(c;T)>IC(g(c))*) for any *c* in *C*.

In practice, the *IC* of *c* is computed as the inverse of its probability of occurrence:

$$IC(c) = -\log p(c). \tag{1}$$

The probabilities needed to measure the informativeness of textual terms are usually assessed from corpora [39]. In this way, general terms will provide less information than more specific ones because the probability of appearance in a discourse of the former is greater.

Likewise, the *PMI(c;T)* is computed as the normalized probability of co-occurrence of *c* and $T=\{t_1,...,t_n\}$, given their joint and marginal probabilities:

$$PMI(c;T) = \log \frac{p(c,t_1,...,t_n)}{p(c)p(t_1,...,t_n)}. \tag{2}$$

To capture a realistic notion of disclosure, probabilities of (co-)occurrence should be assessed from a corpus that truthfully represents the knowledge *K* available to potential attackers to perform their inferences. Thus, in the worst-case scenario (i.e., perfectly knowledgeable attackers), the corpus should be large and heterogeneous enough to cover and capture the information distribution at a social scale. As discussed in [20], the Web is well-suited for this purpose, since it offers the largest amount of directly accessible information/knowledge sources and it is so large and heterogeneous that it can be considered a realistic proxy for social knowledge. Moreover, term probabilities can be computed efficiently by querying terms and term combinations in publicly available Web search engines and evaluating the resulting page count [40] (thus avoiding the need to locally store a large textual corpus for IC calculation):

$$p(t) = \frac{web\_page\_count(t)}{total\_webs}, \tag{3}$$

where *total_webs* is the number of webs resources indexed by the Web search engine.

We deployed this Web-based information theoretic assessment of disclosure risks to detect risky terms at the proxy installed at the user side. In the following we detail the algorithm that implements this process.

---

**Algorithm 1. Detection of risky terms**

```
Input:   D    //the input document
         C    //the ordered set of entities to be protected
         g(C) //the ordered set of generalizations corresponding to C (optional)
Output:  RT   //the set of risky terms

01  for (n=1;n<=MAX;n++) do //cardinality of the combination of terms to evaluate
    //create all combinations of terms in D of cardinality n sorted by informativeness
02      Comb_n=createCombinations(D,n);
03      while (not(empty(Comb_n))) do //evaluate each combination
04          T=getCombination(Comb_n); //obtain the next (set of) term(s)
    //evaluate if PMI(c_k;T)>IC(g(c_k)) ∀ c_k e C until true or c_k=null
05          if (checkDisclosure(C,g(C),T) then
06              add(T,RT); //add the risky term(s) T to RT
07              remove(T,Comb_n,D); //remove T in the combinations and in D
08          end if
09      end while
10  end for
11  return RT;
```

---

Algorithm 1 iteratively generates and evaluates all possible combinations of terms (co-) occurring in *D* with increasing cardinality (lines 1 and 2). Thus, in the first iteration, the elements of *Comb₁* are single terms, which are evaluated for disclosure (line 5), whereas in subsequent iterations, we evaluate the risk caused by the co-occurrence of *several* terms. In lines 5-6, those terms *t* (or group of terms *T*) in the input document whose *PMI(c;t)>IC(c)* (or *PMI(c;T)>IC(c)*) for any *c* in *C* are tagged risky. If any of these terms are risky according to the privacy criterion stated by the model instantiation, they are added to the set of risky terms *RT* (line 6); moreover, they are also removed from the remaining combinations of the same cardinality (in *Comb_n*) and also from the input document *D*, so that they are no longer considered when creating new combinations (line 2) with larger cardinality. By iteratively removing terms already assessed as risky, we minimize the number of combinations (and, thus, of queries to the Web Search Engine) that should be evaluated for disclosure.

## 4.2. Semantic data splitting and distributed storage

In the standard context of document redaction/sanitization, in which protected versions of the input documents should be released atomically, terms detected as risky are usually removed or generalized, so that the level of disclosure they caused is lowered below that specified in the privacy requirements. This produces a loss of utility, since some of the original terms are no longer present (in their fully specified form) in the protected output.

In a multi-cloud storage scenario, however, we can take advantage of the availability of several (independent) cloud storage locations to split and store *clear* data in a privacy preserving way while retaining –most of- the outsourced functionalities. As introduced in section 2, because the different cloud storage locations are aware neither of each other nor of the splitting criterion implemented by the local proxy, CSPs would not able to reconstruct the complete data in an unequivocal way, thus preventing disclosure. Moreover, since a single cloud location can be shared by several users, because the distributed cloud storage is transparently managed by the proxy, the split clear data of each user will also be merged with that of other users; this adds randomness to the outsourced data stored at each location, and offers a better protection against collusion attacks of several CSPs that may try to aggregate partial (merged) data chunks.

In any case, and in coherence with the *C*-sanitization privacy model on which we rely, the splitting process should guarantee that each separately stored chunk of data fulfills Definitions 1-2 through their information theoretic enforcement. According to the risky terms *RT* detected in the previous step, we may distinguish two cases. First, we find individual terms *t* whose sole occurrence in the document causes a disclosure risk according to the privacy criterion. For example, if we want a medical record to be *cancer*-sanitized, the sole occurrence in the document of synonyms of *cancer* such as *malignant neoplasm* (whose semantics are equal to those of *cancer*), or of specializations such as *breast cancer* or *melanoma* (whose semantics encompass and extend those of their generalization *cancer*), completely discloses the semantics of *cancer*. Figure 5 illustrates both situations in terms of informativeness.

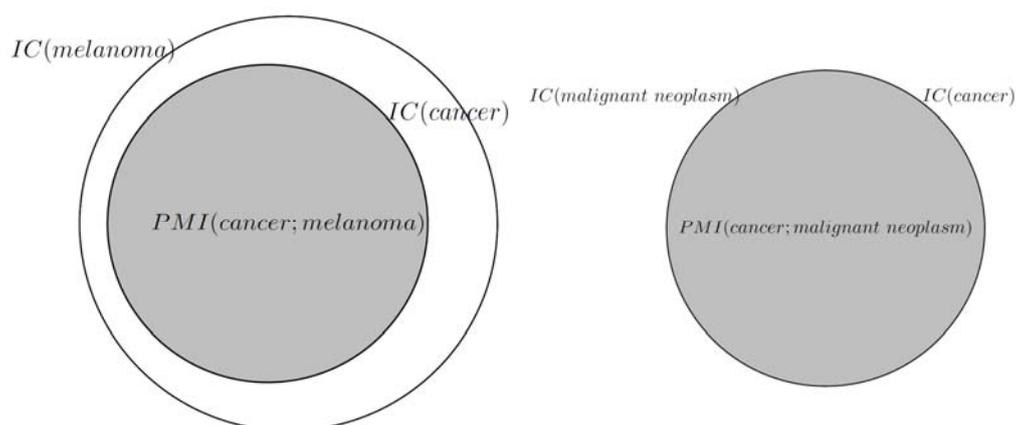

*Figure 5.* Grayed area: disclosure of *cancer*; on the left: *melanoma* is a taxonomical specialization of *cancer*; on the right: *malignant neoplasm* is a synonym of *cancer*.

Moreover, if we use generalizations of *cancer* (e.g., *disease*) to set a stricter privacy criterion in the model instantiation, other individual terms may also cause disclosure. For example, in a (*disease, cancer*)-sanitized document, any specialization of *disease* that is also a generalization of *cancer*, such as *tumor*, will also cause disclosure. These individual terms can be considered *identifiers* of the entity to be protected and, thus, they should not be stored in clear form in *any* cloud location. These terms, which, as it will be discussed latter, represent a small percentage of the amount of terms that should be protected, are stored locally by the proxy.

Second, and more usually, we can identify combinations of terms $RT_i=\{t_1,...,t_j,...,t_n\}$ co-occurring in the document that, even though each $t_j$ does not produce enough disclosure to be considered risky, their aggregation does; that is, they can be considered *quasi-identifiers* of the entity to be protected. In this case, the actual disclosure produced by $RT_i$ is the *union* of the individual disclosures caused by each $t_j$ in $RT_i$ with regard to $c$, which can be computed using Eq. (2). For example, as illustrated in Figure 6, if a medical record should be *cancer*-sanitized, the co-occurrence of the set of terms $RT=\{$*blood in the urine*, *lump, fatigue*$\}$ may be risky because their aggregation is highly correlated with *cancer*, even though each individual term only produces a partial disclosure.

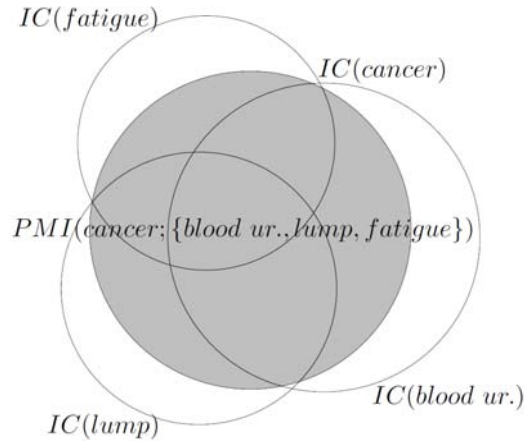

*Figure 6.* Grayed area: disclosure of *cancer* caused by the co-occurrence of *fatigue, lump* and *blood in the urine*.

To protect these combinations of risky terms, we exploit the independent and distributed storage offered by a multi-cloud by separately storing, in clear form, individual terms $t_j$ or subsets of $RT_i$. In this manner, we prevent aggregate inferences and avoid disclosure but retain functionalities, because partial data is stored in clear. The data chunks should be created in a way that, individually, they fulfill the condition stated by the instantiation of the *C*-sanitization model; then, they should be stored separately so that they remain unlinkable to the CSPs. This will also protect against accidental leaks of information by the CSP or external attacks, because only partial (*C*-sanitized) information can be gathered. The remaining terms in the document that were not considered risky can be put together in a separate chunk and stored in a different cloud location.

To reconstruct the whole document and to provide coherent and complete results to user queries, the proxy will store, on local premises, the following metadata for each outsourced document:

- The list of *identifying* terms, which are not outsourced to the cloud due to their individual sensitiveness, sorted according their appearance in the document.
- A list with the cloud locations of the *quasi-identifying* terms $t_j$ in $RT_i$, with their offset within the data chunk in which they are contained. The list is sorted according to the order of appearance of *quasi-identifying* terms in the document.
- The cloud location of the sanitized document resulting from removing *identifying* and *quasi-identifying* terms. The positions they occupied within the document are tagged with sentinels ($id$, $qi$, respectively); in this way, the proxy will be able to reconstruct the document content by simply replacing them with the elements in the local sorted list of identifiers, and the outsourced quasi-identifiers retrieved from the appropriate CSPs, according to the local sorted list of cloud locations and their offsets within such locations.

More details on how data retrieval, search queries and other cloud functionalities exploit these metadata are provided in section 4.4.

The main burden of the data outsourcing process is the number of required locations; also, a large number of locations means that a large number of queries should be performed to gather the whole document content. Two aspects have a main influence on the number of required locations. On the one hand, the instantiation of the privacy model (by means of disclosure thresholds $g(C)$ for the sensitive entities $C$), enables to configure the trade-off between the level of protection and the number of different cloud locations required to fulfill the privacy guarantee; that is, the more general the thresholds, the stricter the protection and the larger the number of locations needed to fulfill the privacy criterion. On the other hand, if documents are large, their discourses are tight (so that co-occurring terms are highly correlated), and/or the privacy requirements are strict, the number of quasi-identifiers to be separately stored could be large. Notice, however, that a cloud location can be shared for different documents and users.

To minimize this cost, and to render the solution practical, in the following we propose a greedy splitting and allocation algorithm that incorporates several heuristics to minimize the number of cloud locations *cLoc* required to store documents while ensuring that each data chunk fulfills the privacy model. The main idea of the algorithm is to put as many terms $t_j$ in each chunk, *chunk$_p$*, as possible, while their aggregation still fulfills the privacy criterion (i.e., the disclosure allocation "budget" of the chunks has not been consumed); only when the criterion is not fulfilled, a new chunk is created and the process is repeated again until no more terms remain to be allocated.

---

### Algorithm 2. Data splitting and distributed storage

```
Input:  D    //the input document
        RT   //the set of risky terms
        C    //the ordered set of entities to be protected
        g(C) //the ordered set of generalizations corresponding to C (optional)

01 D'=D; //D' will be the sanitized version of D
02 while ((not(empty(RT)))) do
03    RTᵢ=getSensitiveCombination(RT);
04    if (|RTᵢ|==1) then //if RTᵢ is an individual term: an identifier
05       replace(RTᵢ,"$id$",D'); //replace RTᵢ by the string $id$ in D'
06       add(RTᵢ,list_ids); // the term RTᵢ is added to a list of identifying terms
07    else //RTᵢ is a set of terms
08       replace(RTᵢ,"$qid$",D');
09       ITᵢ=getMostInformativeTerm(RTᵢ); //allocate first the term with highest IC
10       while (not(empty(RTᵢ)) //until all terms in RTᵢ have been allocated
11          location_found=false;
12          chunkₚ=first(Chunk_set); //a candidate chunk to store ITᵢ
13          while (not(location_found) && chunkₚ≠null) do
14             T=chunkₚ+ITᵢ; //concatenate all terms in chunkₚ with ITᵢ
15             if (not(checkDisclosure(C,g(C),T))) then  //PMI(cₖ;T)≤IC(g(cₖ)) ∀cₖ ϵ C
16                location_found=true; //ITᵢ does not cause disclosure in chunkₚ
17                add(ITᵢ,chunkₚ);
18                remove(ITᵢ,RTᵢ); //ITᵢ has been already allocated
19                ITᵢ=getRemainingTerms(RTᵢ); //evaluate the remaining terms
20             else
21                chunkₚ=next(Chunk_set); //if disclosure, check the next chunk
22             end if
23          end while
24          if (not(location_found)) then
```

```
25            if (|IT₁|==1) then  //IT₁ cannot be split
26                chunkₚ=new_chunk(IT₁);  //create a new chunk with IT₁
                  //add the new chunk sorted by increasing disclosure level
27                addSorted(chunkₚ,Chunk_set);
28                remove(IT₁,RT₁);  //IT₁ has been already allocated in a new chunk
29                IT₁=getRemainingTerms(RT₁);
30            else  //if IT₁ can be split
31                IT₁=getMostInformativeTerm(RT₁);
32            end if
33          end if
34        end while
35    end if
36  end while
37  for each chunkₚ in Chunk_set do
38      cLoc=storeSeparateCloudLocation(chunkₚ);  //outsource the chunk to the cloud
39      add(cLoc,list_cLoc_chunks);  //add it to a list of cloud locations
40  end for
40  doc_cLoc=storeSeparateCloudLocation(D');  //outsource the sanitized document
41  storeLocally(doc_cLoc);  //store locally the location of the sanitized doc
      //generate the sort lists of ids and of chunk locations+offsets for quasi-ids
42  [sort_list_ids,sort_list_cLoc]=getSortList(list_ids,Chunk_set,list_cLoc_chunks,D);
43  storeMetadataLocally(sort_list_ids,sort_list_cLoc);
```

Algorithm 2 iteratively allocates each set of risky terms $RT_i$ detected by Algorithm 1. If terms are individually risky, those replaced by the sentinel \$id\$ within the sanitized version $D'$ of the original document $D$ (line 5) are temporally stored. If we are dealing with a *set* of quasi-identifying terms, we replace each of them by the sentinel \$qid\$ in $D'$ (line 8) and we first try to allocate the term $IT_i$ within the set $RT_i$ with the highest information content, because it is the one that would likely cause the largest disclosure. The algorithm then iterates the set of data chunks that have been created so far (lines 13-23) and evaluates whether the terms already in the *chunk_p* plus the new term to be allocated ($IT_i$) fulfill the privacy criterion (line 15). If this is the case, $IT_i$ is stored in the chunk and it is removed from the set of risky terms $RT_i$ to which it belonged (lines 18-19). Since we have removed $IT_i$ from $RT_i$, by definition, the remaining terms in $RT_i$ would not cause disclosure (i.e., a set of quasi-identifiers that only cause a risk when all of them are evaluated in aggregate); thus, in the next iteration, we try to allocate all the remaining terms in $RT_i$ at once (line 19) in order to minimize the allocation effort that an individual allocation of terms would represent (mainly, in terms of web queries).

If none of the available chunks have enough allocation budget (or we are at the beginning of the process and we have not created any chunk yet), a new chunk containing only the most informative term $IT_i$ (line 26) is created, or we iteratively split and evaluate the remaining terms in $RT_i$, starting by the most informative one (line 31).

The iterative process continues until all the terms in the risky combination $RT_i$ have been allocated in a chunk (line 10) and all risky combinations in $RT$ have been considered (line 2). Then, each chunk with quasi-identifiers is outsourced to a separate cloud location *cLoc* (lines 37-40), whereas the sanitized document $D'$ containing non-risky terms is outsourced to a single cloud location, which is locally stored (lines 40-41). Finally, the list of *identifiers* terms (*sort_list_ids*) and the list of cloud locations for the outsourced *quasi-identifying* terms, plus their offsets within the data chunk in which they are contained (*sort_list_cLoc*), are created and locally stored as metadata (lines 42-43). As detailed above, these metadata are needed to locally reconstruct the document $D$ from its sanitized version $D'$ (retrieved from the cloud), as we will detail in section 4.4; to make this possible, their contents (i.e., identifying terms and cloud locations of quasi-identifying terms, respectively) are sorted in the order of appearance of the terms in the original document $D$.

### 4.3. Minimizing the number of cloud locations

To minimize the number of cloud locations required to store data while fulfilling the privacy criterion, Algorithms 1 and 2 incorporate several heuristics to prune and guide the evaluation of risky term sets and data chunks. Notice that the optimal minimization (with regard to the assessment of combinations and the allocation of their terms) is NP-hard, because it would require evaluating all the possible combinations of any cardinally and every possible allocation of terms; this hardness, which is common to any privacy method that aims at balancing the trade-off between privacy and data utility [41], is impractical for large documents and would require an enormous number of probability calculation queries to evaluate the informativeness of all possible combinations of any cardinality. To render the solution practical, the greedy algorithms detailed above are guided by the following heuristics:

- Evaluate first the disclosure risk of combinations of terms $T$ with the lowest cardinality (see Algorithm 1). In this manner, we avoid starting the analysis with large sets of terms for which, in case of disclosure of a certain $c_k$, we would not able to discern which of the $t_j$ in $T$ are strongly related with $c_k$ and, thus, effectively cause the disclosure, and those that are irrelevant. Thus, to avoid unnecessary splitting, which will require a larger number of locations to accommodate the terms to be protected, we begin the analysis with individual terms, and only when the disclosure of individual terms has been assessed, we move to combinations of cardinality 2, 3, etc. Moreover, once an individual term or a combination of terms is identified as risky, those terms will not be considered in the next iteration (line 7 of Algorithm 1), either during the evaluation of combinations of the same cardinality, or during the creation of subsequent combinations with larger cardinality. In this manner, we prune unnecessary assessments of disclosure risk.

- At each iteration, both algorithms evaluate first those terms with the highest informativeness. In Algorithm 1, this is applied when creating term combinations (line 2). The idea is to begin the assessment of disclosure with the terms that, due to their inherently high informativeness, are more likely to reveal the semantics of the entities in $C$. In this manner, the most potentially risky terms (and their combinations) are evaluated first, thus contributing to minimize the amount of terms tagged as risky, because risky terms cannot be used anymore to create new combinations (line 7 of Algorithm 1). Likewise, in Algorithm 2, the informativeness of terms is used to guide their allocation (line 9 of Algorithm 2). Highly informative risky terms are the first that are allocated to chunks, which will likely correspond to an empty chunk, because these are the ones that impose the stricter restriction and, thus, the most difficult to allocate. This strategy will also minimize the number of unsuccessful evaluations of disclosure during the process of location search.

- Finally, chunks are sorted and evaluated according to the level of aggregated disclosure that the terms already allocated to them produce (line 27 of Algorithm 2). The chunk with the lowest disclosure is evaluated first when allocating a new term. With this strategy, we try to maximize the chance of discovering a valid chunk with the least number of attempts (because we try the easiest allocation first), and tend to produce a more even distribution of terms in the different chunks, so that chunks incur in similar levels of disclosure.

By using the heuristic greedy Algorithms 1 and 2, we transform a NP-hard problem into a feasible one that scales quadratically w.r.t. the number of terms in the document; this cost is in line with other data protection mechanisms that aim at balancing the tradeoff between privacy protection and data utility [42]. Moreover, the heuristics we use to bind, sort and prune the evaluation of terms contribute significantly lower the number of actual combinations to evaluate.

### 4.4. Queries and data retrieval

One of the main benefits of data splitting over cryptographic approaches is the possibility to straightforwardly support functionalities over outsourced split data. In this section, we discuss how user queries requesting functionalities over the documents split with our method can be satisfied by the local proxy.

- *Document retrieval*: to retrieve the whole content of a specific document, the proxy will retrieve from its local metadata database i) the sorted list of identifying terms (*sort_list_ids*) ii) the sorted list of cloud locations plus the offset within the data chunk of the quasi-identifying terms (*sort_list_cLoc*) and iii) the location of the outsourced sanitized document *D'* (*doc_cLoc*). Then, as many retrieval queries as document chunks (plus one more for the sanitized document *D'*) will be executed in parallel toward the respective cloud locations. Once all partial results have been gathered, the proxy will reconstruct the complete document by i) replacing removed identifiers ($id$) in the sanitized document with those locally stored in the same order and, ii) replacing removed quasi-identifiers ($qi$) in the sanitized document *D'* with those retrieved from the clouds, in the same order, and by considering their relative offset within their chunk.

- *Keyword searches*: complex keyword searches involving one or several keywords even involving regular expressions can be easily implemented by broadcasting the search query over all the pieces, including the locally stored identifiers and those pieces outsourced to the cloud (as stated in the metadata), of a specific document or a collection of documents. Then, the set of Boolean results (i.e., query match/mismatch) obtained for each document can be locally aggregated with an OR operator in order to obtain the complete query result (i.e., whether the document, as a whole, matches the query). For conjunctive queries involving Boolean combinations of keywords (AND, OR, NOT), a similar process can be implemented by i) decomposing the conjunctive query in several non-conjunctive queries (one for each keyword), ii) broadcasting the queries over the different document pieces, and iii) aggregating the set of partial results of each non-conjunctive query with an OR and iv) aggregating the aggregated non-conjunctive query results with the initial Boolean combination. If the search query also expects retrieving the matching document(s), the previous process will be followed by the document retrieval depicted above.

- *Document updates*: partial updates corresponding to modifications performed over a retrieved document are also straightforward. For partial deletions, removed terms only need to be removed either from the local metadata database (if they were identifiers) or from the outsourced document chunks. Because removing terms from outsourced chunks lowers the amount of disclosed information per chunk, it does not violate the privacy requirements. For term replacements, we only need to check whether the term replacement fulfills the privacy requirements in the document chunk of the original term: in the affirmative case, we can just replace the term in the chunk; otherwise, we need to remove the original term and allocate the replacement in a chunk suitable for the privacy requirements. Term additions are handled similarly to the latter case: the proxy evaluates their sensitiveness and allocate them in an existing or new chunk, accordingly. In all cases, the local metadata is updated to properly reconstruct the document.

## 5. Security and functional analysis

In this section, we discuss the security of the splitting mechanism in comparison with cryptographic approaches. Moreover, we also discuss the preservation of outsourced functionalities and the effort needed to support them with the two approaches.

With data splitting, each cloud receives a data chunk that fulfills the privacy requirements. Since these requirements are defined as an instantiation of the *C*-sanitization privacy model, we have the a priori privacy guarantee that the level of disclosure of each data chunk does not disclose more sensitive information than that defined in the privacy requirements (the disclosure threshold $g(c)$) for each sensitive entity ($c$) (Definition 2). This a priori guarantee is, in essence, conceptually equivalent to that of other well-known privacy models for database releases [43] (such as $k$-anonymity [27, 28]): with $C$-sanitization-driven data splitting we ensure that potential disclosure inferences are ambiguous (i.e., non-unequivocal) for each sensitive entity $c$, with a level of ambiguity defined by the disclosure threshold. Obviously, because split data are stored in clear, some partial information is leaked to each cloud (as it is the case for $k$-anonymous data releases that also assume that the released data should be useful for third parties), but with the guarantee that $c$ cannot be unequivocally disclosed (also as with $k$-anonymity, which ensures that the probability of disclosing the protected entity is $1/k$).

Assuming that clouds do not collude, further operations over the split outsourced data, including retrieval, queries, updates, and deletions, which are independently executed over individual cloud locations, would reveal no additional information to each individual cloud. If the clouds collude, they may try to reconstruct data by aggregating data chunks, but this can be hardly made unequivocally because of the potential number of feasible aggregations without knowing the splitting criterion. Moreover, the fact that a single cloud location may indistinctly store partial data of several independent documents (i.e., *merged* split data) will largely increase the number of feasible aggregations and, thus, decrease the probability of a successful disclosure inference [44].

Other than the evaluation of the sensitiveness of terms (which is done once at the storage stage), no other costly operation is needed to manage and query data, which are functionalities much more common than storage, and for which fast response times are expected. Moreover, as detailed in section 4.4, search, retrieval and update queries with a high level of expressiveness (e.g., conjunctive queries, regular expressions, etc.) can be seamlessly supported by broadcasting the query to the appropriate cloud locations and aggregating the results in coherence with the query syntax. Moreover, partial queries are transparently and efficiently executed by cloud service providers, which do not need to be aware of the fact that data are split (i.e., they do not need to modify the logic of their services); cloud providers also receive clear partial data that are still analytically useful. Finally, local aggregations made on the proxy side on partial results are also efficient, since they consist on logical operators on clear data. As a matter of fact, all the operations on outsourced data discussed in the former section scale linearly to the number of terms in the document, in the worst case.

Moreover, the multi-cloud notion exploited by our data splitting mechanism may not only be used to enforce privacy-preserving storage, but to ensure robustness against cloud failures: by redundantly storing the data chunks in different cloud locations, we ensure the availability of the data in the event of failure or data loss by one or several CSPs. Obviously, additional cloud locations would be needed to redundantly store data chunks, in the same way as additional physical drives are needed to achieve data redundancy on local premises.

In comparison, functionality-preserving encryption mechanisms, which can be either based on homomorphic encryption or on searchable encryption, offer more robust privacy because no clear data is outsourced to the cloud, but they require managing sensitive cryptographic material. Moreover, the fact that the complete data is stored in a unique cloud makes it possible for the cloud provider to learn some query and data patterns, such as which pieces of data seem to be correlated because they are queried conjunctively. From a functional perspective, they also suffer from several limitations. On the one hand, homomorphic encryption [6] allows processing encrypted data without decrypting them, but the computational cost is prohibitive, both for clouds providers, which execute the operations, and for the local proxy, which orchestrates the process and decrypts the results. Moreover, in order to coherently operate over the encrypted outsourced data, so that a correct result can be decrypted on local premises, the cloud provider should use specially tailored operators. Thus, cloud providers should be aware of the fact that they are dealing with encrypted data and should agree on modifying the logic of their services. This is unfeasible in many commercial scenarios, especially for free cloud services, for which cloud providers expect gaining some utility from the users' data.

On the other hand, searchable encryption [7] offers a more computationally viable solution to support searches and retrieval on encrypted data, but still requires adapting the cloud services to operate on such encrypted data. Moreover, additional data (i.e., encrypted search indexes) need to be attached to the outsourced encrypted dataset and, because of this, all the queries that are expected to be performed over the data should be known beforehand. This complicates data updates, which usually require retrieving, decrypting and updating the whole dataset, re-creating the search indexes, and re-encrypt and re-upload the dataset to the cloud. Moreover, computationally complex solutions (both for clouds and for the local proxy) relying on additional search indexes (e.g., hierarchical indexes) and multiple successive queries are needed to support queries other than single keyword searches [45]. Finally, even though some approaches have been proposed to support queries with multiple keywords [46] or conjunctive queries [47], they are still far from the fully transparent, efficient and flexible query support offered by data splitting.

# 6.  Evaluation

There are two aspects of the proposed system that can be evaluated: the accuracy of the detection of risky terms (which is based on the information theoretic enforcement of the *C*-sanitization model) and the effectiveness of the splitting process. The results of the former have been already reported, evaluated and discussed in [25, 26], which shows that the implementation of the *C*-sanitization model with moderately general thresholds achieves a performance comparable to that of human sanitizers. In the following, we report the results and evaluate the effectiveness of the splitting process and the proposed heuristics in reducing the number of cloud locations with a corpus similar to that used in the literature on document sanitization [24, 25, 48].

## 6.1.  Evaluation data and case studies

As evaluation data, we took the same document corpus that was used in [20, 25, 26] to measure the accuracy of the detection of risky terms. It consists of a set of Wikipedia articles describing some entities/topics that are sensitive according to the current legal frameworks on data privacy (HIV, Sexually Transmitted Diseases (STDs), homosexuality and Catholicism) and a highly sensitive individual (Osama Bin Laden). Regarding the former, U.S. federal laws on medical data privacy [35, 36] mandate hospitals and healthcare organizations to protect any references made to STDs and HIV status in patient medical records before releasing them to, for example, insurance companies in response to Worker's Compensation or Motor Vehicle Accident claims. To do so, those terms explicitly referring to these diseases and those semantically related ones, such as treatments or symptoms, should be protected [19]. Likewise, the EU Data Protection Directive [38] states that the information related to the religion and sexual orientation of EU citizens should be protected in order to avoid possible discrimination. In all these cases, *attribute disclosure* protection should be provided (i.e., the confidential information related to the individual should not be disclosed). For the case of Osama Bin Laden, the identity of the individual should not be disclosed [20], that is, the document should protect against *identity disclosure*. Wikipedia articles have been used by related works since they configure a challenging data protection scenario because of their tight and semantically rich discourses [20].

To offer the kind of data protection needed for each document, we have employed different instantiations of the *C*-sanitization model. First, a basic *HIV- STD- Homosexuality-Catholicism- Bin Laden*-sanitization has been individually applied for the respective documents, in which individual terms or term combinations that unequivocally reveal the entities to be protected were considered risky. Afterwards, semantically coherent generalizations (extracted from WordNet) have been used as thresholds to state the maximum level of allowed disclosure. *Virus*, *Infection* and *Condition* were used as generalizations of *HIV*, *Disease* for *STD*, *Sexuality* for *Homosexuality*, *Religion* for *Catholicism* and *Person* for *Bin Laden*.

These case studies also illustrate how the *C*-sanitization model can be intuitively instantiated in coherence with the privacy notions stated in current legislations on data protection (which are defined at a conceptual -rather than statistical- level) and for different privacy requirements (i.e., different levels of protection against *identify* and/or *attribute* disclosure).

## 6.2. Testing configurations and evaluation metrics

To evaluate the benefits brought by the semantic data splitting algorithm, the following splitting strategies have been implemented and applied for the evaluation corpus:

- A naïve approach in which each sensitive term is stored in an individual location. This completely uninformed strategy trivially fulfills the privacy requirements, even though it would likely require a large number of locations.
- Algorithm 2 without the heuristics that sort the terms and chunks in order to maximize the chance of finding a valid allocation. In this case, terms are managed in the same order as they appear in the document and chunks are evaluated by their order of creation.
- Algorithm 2 with the proposed heuristics, as detailed in Section 4.2.

To measure the effectiveness of each splitting strategy, we defined and quantified the following quality metrics:

- Number of cloud locations needed to ensure that all the data chunks stored in clear form fulfill the privacy requirements. This is a crucial aspect because, in general, the larger the number of locations is, the higher the storage cost the user and/or the CSP will incur in. In addition, more locations imply that more (partial) queries should be broadcasted to fulfill a given query and more (partial) results should be aggregated to provide the complete result to the user (as detailed in section 4.4). Notice that, in all cases, an additional location will be needed to store the remaining non-risky text from the input document.
- Disclosure balance between cloud locations. Because sets of terms are stored together according to the aggregated amount of information they disclose, a good allocation (in terms of an optimal use of the available resources) will be such that the aggregated disclosure at each location is, in average, the nearest (but always below) to the disclosure threshold defined by the model instantiation. Moreover, a good disclosure balance will be reflected by a low variance between the disclosure level of the term sets stored at each location; in this way, we achieve a similar level of (controlled) disclosure and avoid that some locations disclose significantly more information whereas others disclose significantly less. Specifically, the heuristics presented in section 4.3, which sort locations and terms to allocate, pursue this even balancing. To measure the effectiveness of the heuristics, we also computed the arithmetic average and the standard deviation of the aggregated disclosure of the set to terms stored in a chunk at each location. To offer figures that can be directly compared for the different documents and model instantiations, we normalized each of these metrics by the informativeness of the disclosure threshold ($g(C)$) stated in the model instantiation:

$$Normalized\_metric = \frac{metric}{IC(disclosure\_threshold)} \times 100, \qquad (4)$$

where *metric* corresponds to either the arithmetic average or the standard deviation of the aggregated disclosure of the locations in use (computed with Eq. (2)). A good normalized average will be such that is near 100% (because it used most of the allocation budget of the locations in use), whereas a good normalized deviation will be such that is near 0% (because all the locations offer a similar level of disclosure).

## 6.3. Results and discussion

The evaluation metrics for the different documents, model instantiations, splitting strategies and heuristics in use are shown in Tables 1-4. We also show in these tables the percentage of terms from the whole document that were tagged as *identifiers* (i.e., they individually disclose more information than that allowed by the model instantiation) and *quasi-identifiers* (i.e., those that only cause disclosure when they co-occur with other quasi-identifiers).

*Table 1.* Evaluation metrics for the document about *HIV* with different splitting strategies and privacy requirements.

| Privacy model instantiation | % Identifiers | % Quasi-identifiers | Splitting strategy | #Locations (Q-Ids) | Norm. Avg. Disclosure | Norm. Std. Dev. |
|---|---|---|---|---|---|---|
| *HIV*-sanitization | 6.9% | 15.2% | Naïve | 24 | 76.6% | 20.01% |
| | | | No heuristics | 10 | 89.35% | 17.60% |
| | | | Heuristics | 9 | 90.20% | 9.76% |
| (*HIV, Virus*)-Sanitization | 10.1% | 13.9% | Naïve | 22 | 79.7% | 14.86% |
| | | | No heuristics | 10 | 91.0% | 7.46% |
| | | | Heuristics | 9 | 94.87% | 3.54% |
| (*HIV, Infection*)-sanitization | 6.3% | 10.8% | Naïve | 17 | 78.24% | 24.4% |
| | | | No heuristics | 9 | 95.42% | 6.81% |
| | | | Heuristics | 6 | 94.55% | 5.67% |
| (*HIV, condition*)-sanitization | 23.4% | 1.3% | Naïve | 2 | 78.91% | 3.03% |
| | | | No heuristics | 2 | 78.91% | 3.03% |
| | | | Heuristics | 2 | 78.91% | 3.03% |

*Table 2.* Evaluation metrics for the document about *STDs* with different splitting strategies and privacy requirements.

| Privacy model instantiation | % Identifiers | % Quasi-identifiers | Splitting strategy | #Locations (Q-Ids) | Norm. Avg. Disclosure | Norm. Std. Dev. |
|---|---|---|---|---|---|---|
| *STD*-sanitization | 5.7% | 10.7% | Naïve | 56 | 65.06% | 39.28% |
| | | | No heuristics | 19 | 78.65% | 22.52% |
| | | | Heuristics | 15 | 86.87% | 16.59% |
| (*STD, disease*)-sanitization | 4.4% | 11.8% | Naïve | 62 | 65.84% | 34.7% |
| | | | No heuristics | 26 | 74.7% | 23.39% |
| | | | Heuristics | 19 | 90.92% | 20.48% |

*Table 3.* Evaluation metrics for the document about *Catholicism* with different splitting strategies and privacy requirements.

| Privacy model instantiation | % Identifiers | % Quasi-identifiers | Splitting strategy | #Locations (Q-Ids) | Norm. Avg. Disclosure | Norm. Std. Dev. |
|---|---|---|---|---|---|---|
| *Catholicism*-sanitization | 4.8% | 14.5% | Naïve | 18 | 77.52% | 26.44% |
| | | | No heuristics | 8 | 84.59% | 25.05% |
| | | | Heuristics | 7 | 87.14% | 15.3% |
| (*Catholicism, religion*)-sanitization | 9.7% | 9.7% | Naïve | 12 | 79.21% | 21.42% |
| | | | No heuristics | 7 | 80.82% | 12.4% |
| | | | Heuristics | 6 | 86.49% | 7.32% |

*Table 4.* Evaluation metrics for the document about *homosexuality* with different splitting strategies and privacy requirements.

| Privacy model instantiation | % Identifiers | % Quasi-identifiers | Splitting strategy | #Locations (Q-Ids) | Norm. Avg. Disclosure | Norm. Std. Dev. |
|---|---|---|---|---|---|---|
| *homosexuality*-sanitization | 2.1% | 12.9% | Naïve | 48 | 63.04% | 51.21% |
| | | | No heuristics | 8 | 77.31% | 28.1% |
| | | | Heuristics | 5 | 81.72% | 15.78% |
| (*homosexuality, sexuality*)-sanitization | 2.4% | 13.8% | Naïve | 50 | 63.81% | 40.02% |
| | | | No heuristics | 15 | 75.29% | 21.98% |
| | | | Heuristics | 13 | 81.79% | 17.36% |

*Table 5.* Evaluation metrics for the document about *Osama Bin Laden* with different splitting strategies and privacy requirements.

| Privacy model instantiation | % Identifiers | % Quasi-identifiers | Splitting strategy | #Locations (Q-Ids) | Norm. Avg. Disclosure | Norm. Std. Dev. |
|---|---|---|---|---|---|---|
| Bin Laden-sanitization | 0.3% | 7.2% | Naïve | 20 | 64.54% | 21.16% |
| | | | No heuristics | 3 | 88.63% | 4.28% |
| | | | Heuristics | 2 | 94.81% | 3.56% |
| (Bin Laden, person)-sanitization | 5.4% | 10.8% | Naïve | 30 | 71.1% | 30% |
| | | | No heuristics | 15 | 71.2% | 24.87% |
| | | | Heuristics | 10 | 73.5% | 20.72% |

First, we observe that the percentage of *identifying* terms increases as the sanitization threshold becomes more general. This is coherent, because a general threshold means that only the most general terms related to entity to be protected can be left in clear, and that any term that is a specialization of the disclosure threshold becomes an *identifier* that cannot be stored in clear without violating the privacy requirements. Notice that, because of the need to locally store them within the proxy, they consume local storage resources. However, one can see from the tables that the locally stored *identifiers* account a small percentage of the total amount of data to be outsourced; moreover, it is important to recall that the kind of documents we are using here constitute a worst-case scenario for data protection: in practice, most document would not be as disclosive and focused as Wikipedia articles describing a sensitive entity.

We also observe that the number of locations needed to store quasi-identifiers in a privacy preserving way also increases as the threshold becomes general, regardless of the splitting strategy. Again, a general threshold means that more and more terms co-occurring in the document can disclose, in aggregate, more information than that specified by the threshold. However, the strictest privacy criterion imposed by the more general thresholds limits the number of terms that can be stored together and, thus, increases the number of locations we need. The only exception is the *HIV* document, for which most quasi-identifiers become identifiers, which can be stored in just one location but in an encrypted form, when setting a very general threshold.

By comparing the different splitting strategies, we clearly see that the naïve approach provides the worst results, since it requires the largest number of locations. Algorithm 2, on the other hand, significantly reduces the number of needed locations since it tries to fit as many terms as possible in each location as long as the privacy requirements are fulfilled. The heuristics incorporated into the algorithm also help to improve the allocation efficiency. By sorting terms in decreasing order of informativeness, we tend to allocate first those terms that impose the strictest restriction and that will consequently be more difficult to allocate. On the other hand, by sorting locations in increasing order of aggregated disclosure, we try first those chunks in which it is more feasible to fit a new term because there is still disclosure budget before violating the privacy requirements. This not only provides a more efficient use of the available resources, but decreases the number of unsuccessful allocation attempts in comparison with the non-heuristic strategy. The heuristic approach provides the largest relative improvement over the other strategies when the least strict privacy criteria are specified, i.e., when no generalizations are used as disclosure thresholds. Indeed, this setting provides more degrees of freedom to allocate terms while still fulfilling the privacy model, which our algorithms takes advantage of.

The effectiveness of the allocation process is also illustrated by the balance of disclosure levels through the different locations in use. We see that, for the heuristic approach, the normalized average is always the highest, and quite near 100%; this indicates that the disclosure budget offered by each location with regards to the threshold has been mostly consumed. Likewise, the normalized deviation between locations is always very low (3-16%), which shows that the disclosure levels of the different locations have been evenly balanced. If we disable the heuristics, results are slightly worse and they get much worse with the naïve approach. In this last case, the allocation is almost random, which means that a lot of allocation budget is wasted.

Results are also coherent for all the documents, regardless that the privacy requirements refer to confidential data (i.e., attribute disclosure), such as diseases, or to identifying data (i.e., identity disclosure), such as person names. Qualitatively, we observed that the heuristic splitting strategy "smartly" tended to store together those terms that offer redundant information (e.g., different lexicalizations or synonyms) because their co-occurrence does not significantly increase the disclosure level of the chunk, whereas disjoint complementary terms with minimal mutual information are stored separately.

## 7.  Conclusions and future work

So far, data encryption has been the main privacy protection technique employed to protect data outsourced to the cloud, even though it usually hampers efficiency, both on cloud and local sides and both for storage and search/retrieval operations, it is not transparent for cloud providers, offers a limited support for outsourced functionalities and adds a number of burdens, such as local management of keys, need to deploy specially tailored software modules in the cloud to support certain outsourced functionalities, etc. In contrast, the semantic data splitting mechanism we propose in this paper offers an interesting alternative: because *all* the outsourced data are stored in (partial) clear form, data management is scalable and efficient, especially during the most common search and retrieval operations, it is fully transparent for cloud providers, which are not aware of the data protection and that can maintain their services intact, and outsourced functionalities are straightforwardly preserved, even for highly expressive search queries. Moreover, our approach builds on the *a priori* privacy guarantees offered by the *C*-sanitization model [25] that, contrary to other more abstract privacy models focused on numerical and structured data, can be intuitively instantiated at a conceptual level by using linguistic labels. This allows users to easily define their privacy requirements without requiring them to be aware of the technicalities related to data protection, and also permits a seamless enforcement of available legislations on data privacy, which are also qualitative. The model also allows defining the degree of protection applied over the data as a function of the amount of semantics that can be disclosed. As it has been shown in the evaluation, this also enables to balance the tradeoff between the level of privacy and the amount of resources (i.e., cloud locations and local storage) needed to enforce it.

Compared to other protection mechanisms based on data splitting, the inherently semantic ground of our approach makes it possible to automatically evaluate the privacy risks of the data according to the semantics they disclose. This relieves the users from the burden of manually identifying sensitive pieces of data (e.g., attributes in a database or specific terms in a textual document), as needed by related works [13, 14, 18], and offers a natural support for unstructured textual data, which can be hardly managed by most of the available privacy protection solutions [25].

In this paper, we have specifically focused on textual documents that constitute a privacy challenge because of their textual content and lack of structure. However, our method can also be applied in structured data bases (i.e., by means of vertical splitting). As future work, we plan to study this possibility by centering the risk assessment on automatically detecting combinations of attribute values that cause disclosure. Moreover, since our method also offers some protection against collusions of several CSPs, by seamlessly merging partial data of several users using the same proxy at the same location, we will also evaluate its effectiveness against standard attacks, and we will study how we can improve it by appropriately merging data chunks at each cloud location.

## Acknowledgements and disclaimer

This work was partly supported by the European Commission (projects H2020-644024 "CLARUS" and H2020-700540 "CANVAS"), by the Spanish Government (projects TIN2014-57364-C2-R "SmartGlacis", TIN2015-70054-REDC "Red de excelencia Consolider ARES" and TIN2016-80250-R "Sec-MCloud") and by the Government of Catalonia under grant 2014 SGR

537. M. Batet is supported by a Postdoctoral grant from Ministry of Economy and Competitiveness (MINECO) (FPDI-2013-16589).

# References

[1] IBM Corporation, The Essential CIO. From www.ibm.com/businesscenter/cpe/download0/218842/2011mmciostudy.pdf, in, 2011.

[2] Cloud Security Alliance, Cloud Usage: Risks and Opportunities Report, in, Sep. 2014.

[3] European Network and Information Security Agency, Cloud Computing. Benefits, risks and recommendations for information security. Revision B, in: L. Dupré, T. Haeberlen (Eds.), Dec. 2012.

[4] E. Ramirez, J. Brill, M.K. Ohlhausen, J.D. Wright, T. McSweeny, Data Brokers: A Call for Transparency and Accountability, in, Federal Trade Commission, May 2014.

[5] R. Battistoni, R. Di Pietro, F. Lombardi, CURE—Towards enforcing a reliable timeline for cloud forensics: Model, architecture, and experiments, Computer Communications, 91-92 (2016) 29-43.

[6] C. Gentry, S. Halevi, N.P. Smart, Fully homomorphic encryption with polylog overhead, in: 31st Annual international conference on Theory and Applications of Cryptographic Techniques (EUROCRYPT'12), Springer-Verlag Berlin, 2012, pp. 465-482.

[7] M. Li, S. Yu, N. Cao, W. Lou, Authorized private keyword search over encrypted data in cloud computing, in: 31st International Conference on Distributed Computing Systems (ICDCS '11), IEEE Computer Society, 2011, pp. 383-392.

[8] J. Li, J. Li, X. Chen, Z. Liu, C. Jia, Privacy-preserving data utilization in hybrid clouds, Future Generation Computer Systems, 30 (2014) 98-106.

[9] C-I. Fan, S-Y. Huang, Controllable privacy preserving search based on symmetric predicate encryption in cloud storage, Future Generation Computer Systems, 29 (2013) 1716-1724.

[10] Y. Chen, R. Sion, On securing untrusted clouds with cryptography, in: 9th annual ACM workshop on Privacy in the electronic society, ACM, 2010, pp. 109-114.

[11] D. Zissis, D. Lekkas, P. Koutsabasis, Cryptographic Dysfunctionality-A Survey on User Perceptions of Digital Certificates, in: C. Georgiadis, H. Jahankhani, E. Pimenidis, R. Bashroush, A. Al-Nemrat (Eds.) Global Security, Safety and Sustainability & e-Democracy, Springer Berlin Heidelberg, 2012, pp. 80-87.

[12] M. Rouse, What is a multi-cloud strategy?, in: SearchCloudApplications. WhatIs.com, 2014.

[13] J.-J. Yang, J.-Q. Li, Y. Niu, A hybrid solution for privacy preserving medical data sharing in the cloud environment, Future Generation Computer Systems, 43-44 (2015) 74-86.

[14] H. Dev, T. Sen, M. Basak, M.E. Ali, An approach to protect the privacy of cloud data from data mining based attacks, in: High Performance Computing, Networking, Storage and Analysis (SCC 2012), IEEE Computer Society, 2012, pp. 1106-1115.

[15] G. Aggarwal, M. Bawa, P. Ganesan, H. Garcia-molina, K. Kenthapadi, R. Motwani, U. Srivastava, D. Thomas, Y. Xu, Two can keep a secret: A distributed architecture for secure database services, in: Second Biennial Conference on Innovative Data Systems Research (CIDR 2005), 2005, pp. 186-199.

[16] Z. Wei, S. Xinwei, Data Privacy Protection Using Multiple Cloud Storages, in: International Conference on Mechatronic Sciences, Electric Engineering and Computer, Shenyang, China, 2013.

[17] V. Ciriani, S. De Capitani di Vimercati, S. Foresti, S. Jajodia, S. Paraboschi, P. Samarati, Fragmentation and Encryption to Enforce Privacy in Data Storage, in: J. Biskup, J. López (Eds.) Computer Security – ESORICS 2007, Springer Berlin Heidelberg, 2007, pp. 171-186.

[18] V. Ganapathy, D. Thomas, T. Feder, H. Garcia-Molina, R. Motwani, Distributing data for secure database services, Transactions on Data Privacy, 5 (2012) 253 - 272.

[19] E. Bier, R. Chow, P. Golle, T. H. King, J. Staddon, The Rules of Redaction: identify, protect, review (and repeat), IEEE Security and Privacy Magazine, 7 (2009) 46-53.

[20] J. Staddon, P. Golle, B. Zimmy, Web-based inference detection, in: 16th USENIX Security Symposium, 2007, pp. Article No. 6.

[21] B. Anandan, C. Clifton, W. Jiang, M. Murugesan, P. Pastrana-Camacho, L.Si, t-plausibility: Generalizing words to desensitize text, Transactions on Data Privacy, 5 (2012) 505-534.

[22] D. Sánchez, M. Batet, A. Viejo, Automatic general-purpose sanitization of textual documents, IEEE Transactions on Information Forensics and Security, 8 (2013) 853-862.

[23] D. Sánchez, M. Batet, A. Viejo, Minimizing the disclosure risk of semantic correlations in document sanitization, Information Sciences, 249 (2013) 110-123.

[24] D. Sánchez, M. Batet, A. Viejo, Utility-Preserving Sanitization of Semantically Correlated Terms in Textual Documents, Information Sciences, 279 (2014) 77-93.

[25] D. Sánchez, M. Batet, C-sanitized: a privacy model for document redaction and sanitization, Journal of the Association for Information Science and Technology, 67 (2016) 148-163.

[26] D. Sánchez, M. Batet, Toward sensitive document release with privacy guarantees, Engineering Applications of Artificial Intelligence, 59 (2017) 23-34.

[27] P. Samarati, L. Sweeney, Protecting privacy when disclosing information: k-anonymity and its enforcement through generalization and suppression, in, SRI International Report, 1998.

[28] P. Samarati, Protecting Respondents' Identities in Microdata Release, IEEE Transactions on Knowledge and Data Engineering, 13 (2001) 1010-1027.

[29] V.T. Chakaravarthy, H. Gupta, P. Roy, M.K. Mohania, Efficient techniques for document sanitization, in: 17th ACM Conference on Information and Knowledge Management (CIKM'08), Napa Valley, California, USA, 2008, pp. 843–852.

[30] C. Dwork, Differential Privacy, in: 33rd International Colloquium ICALP, Springer, Venice, Italy, 2006, pp. 1-12.

[31] A. Viejo, D. Sánchez, Enforcing transparent access to private content in social networks by means of automatic sanitization, Expert Systems with Applications, 62 (2016) 148-160.

[32] M. Imran-Daud, D. Sánchez, A. Viejo, Privacy-driven access control in social networks by means of automatic semantic annotation, Computer Communications, 76 (2016) 12-25.

[33] Department of Health and Human Services, The health insurance portability and accountability act of 1996, in, 2000.

[34] N. Terry, L. Francis, Ensuring the privacy and confidentiality of electronic health records, in: University of Illinois Law Review, 2007, pp. 681-735.

[35] U.S. Department of Health & Human Services, Health Information Privacy, in, 2015.

[36] Department for a Healthy New York, New York State Confidentiality Law, in, 2013.

[37] Legal Information Institute, Privacy Protection For Filings Made with the Court, in, 2013.

[38] The European Parliament and the Council of the EU, Data Protection Directive 95/46/EC, in, 1995.

[39] P. Resnik, Using Information Content to Evalutate Semantic Similarity in a Taxonomy, in: C.S. Mellish (Ed.) 14th International Joint Conference on Artificial Intelligence, IJCAI 1995, Morgan Kaufmann Publishers Inc., Montreal, Quebec, Canada, 1995, pp. 448-453.

[40] P.D. Turney, Mining the Web for Synonyms: PMI-IR versus LSA on TOEFL, in: L. De Raedt, P. Flach (Eds.) 12th European Conference on Machine Learning, ECML 2001, Springer-Verlag, Freiburg, Germany, 2001, pp. 491-502.

[41] A. Hundepool, J. Domingo-Ferrer, L. Franconi, S. Giessing, E.S. Nordholt, K. Spicer, P.P.d. Wolf, Statistical Disclosure Control, Wiley, 2013.

[42] J. Domingo-Ferrer, V. Torra, Ordinal, continuous and heterogeneous k-anonymity through microaggregation, Data Mining and Knowledge Discovery, 11 (2005) 195-212.

[43] J. Domingo-Ferrer, D. Sánchez, J. Soria-Comas, Database Anonymization: Privacy Models, Data Utility, and Microaggregation-based Inter-model Connections, Morgan & Claypool Publishers, 2016.

[44] R. Brinkman, S. Maubach, W. Jonker, A lucky dip as a secure data store, in: Workshop on Information and System Security, 2006.

[45] R. Curtmola, J. Garay, S. Kamara, R. Ostrovsky, Searchable symmetric encryption: Improved definitions and efficient constructions, Journal of Computer Security, 19 (2011). 895–934.

[46] L. Ballard, S. Kamara, F. Monrose, Achieving efficient conjunctive keyword searches over encrypted data, in: Information and Communications Security, Springer, 2005, pp. 414–426.

[47] D. Cash, S. Jarecki, C. Jutla, H. Krawczyk, M. Rosu, M. Steiner, Highly-scalable searchable symmetric encryption with support for boolean queries, in: CRYPTO 2013, 2013, pp. 353–373.

[48] D. Sánchez, M. Batet, A. Viejo, Utility-preserving privacy protection of textual healthcare documents, Journal of Biomedical Informatics, 52 (2014) 189-198.